\documentclass[11pt]{article}
\usepackage{tgpagella}
\usepackage{microtype}
\usepackage{fullpage}
\usepackage[varqu,varl,var0,scaled=0.97]{inconsolata}
\usepackage{amsmath,amssymb,amsfonts,amsthm,mathrsfs,mathtools}
\usepackage{hyperref}
\usepackage{url}
\usepackage{subscript}
\usepackage{graphicx}
\usepackage{xspace}
\usepackage{xcolor}         % colors
\usepackage{mleftright}
\usepackage{bm}
\usepackage{physics}
\usepackage{booktabs}

\newtheorem{informal theorem}[theorem]{Informal Theorem}

\theoremstyle{definition}
\newtheorem{definition}{Definition}[section]

\newcommand{\bR}{\mathbb{R}}

\usepackage{xcolor}
\hypersetup{
    colorlinks,
    linkcolor={red!50!black},
    citecolor={blue!50!black},
    urlcolor={blue!80!black}
}

\usepackage{algorithm}
\usepackage{algpseudocode}
\algrenewcommand\algorithmicrequire{\textbf{Input:}}
\algrenewcommand\algorithmicensure{\textbf{Output:}}

\allowdisplaybreaks

\begin{document}

\title{Investigating Privacy Leakage in Dimensionality Reduction Methods via Reconstruction Attack}

\author{Chayadon Lumbut\thanks{Chiang Mai University, Chiang Mai, Thailand.} \and Donlapark Ponnoprat\footnotemark[1]    \footnote{Corresponding author: \href{mailto:donlapark.p@cmu.ac.th}{donlapark.p@cmu.ac.th}}}

\date{}
\maketitle

\begin{abstract}
This study investigates privacy leakage in dimensionality reduction methods through a novel machine learning-based reconstruction attack. Employing an \emph{informed adversary} threat model, we develop a neural network capable of reconstructing high-dimensional data from low-dimensional embeddings.

We evaluate six popular dimensionality reduction techniques: principal component analysis (PCA), sparse random projection (SRP), multidimensional scaling (MDS), Isomap, t-distributed stochastic neighbor embedding ($t$-SNE), and uniform manifold approximation and projection (UMAP). Using both MNIST and NIH Chest X-ray datasets, we perform a qualitative analysis to identify key factors affecting reconstruction quality. Furthermore, we assess the effectiveness of an additive noise mechanism in mitigating these reconstruction attacks. Our experimental results on both datasets reveal that the attack is effective against deterministic methods (PCA and Isomap). but ineffective against methods that employ random initialization (SRP, MDS, $t$-SNE and UMAP). The experimental results also show that, for PCA and Isomap, our reconstruction network produces higher quality outputs compared to a previously proposed network.

We also study the effect of additive noise mechanism to prevent the reconstruction attack. Our experiment shows that, when adding the images with large noises before performing PCA or Isomap, the attack produced severely distorted reconstructions. In contrast, for the other four methods, the reconstructions still show some recognizable features, though they bear little resemblance to the original images. The code is available at \url{https://github.com/Chayadon/Reconstruction\_attack\_on\_DR}.
\end{abstract}

\section{Introduction}

Machine Learning (ML) models have become essential tools for solving complex real-world problems across various domains, including image processing, natural language processing, and business analytics. However, learning from high-dimensional data can be difficult due to the curse of dimensionality and increased computational requirements. To address these issues, dimensionality reduction methods are employed in order to reduce training costs and improve its efficiency.

Popular dimensionality reduction methods include principal component analysis, $t$-SNE~\cite{JMLR:v9:vandermaaten08a}, and UMAP~\cite{McInnes2018}. These methods aim to reduce data dimensions while preserving global and local properties of the original data, ensuring that relationships between data points in higher dimensions are still reflected in lower-dimensional representations. The information retained from the original data is crucial for effective data analysis and visualization.

\textbf{Motivation.} the widespread use of dimensionality reduction methods on sensitive data has posed privacy risks, particularly in sensitive domains like face recognition~\cite{Lin2020}, genomics~\cite{Platzer2013,Kobak2019,Matthew2020}, and healthcare~\cite{Eshghi2019,Dadu2023}. While low-dimensional representations enable efficient analysis and valuable insights, they also pose significant privacy threats. The core of this risk lies in the information preserved by dimensionality reduction methods, in particular, the relationships between data points. This preservation potentially allows adversaries to gain some information about the original high-dimensional data or determine whether a specific individual's data was included in the training set.  \textbf{As different dimensionality reduction methods have different ways to process the data, it is not trivial to measure the privacy risks from publishing the outputs of these methods. To this end, we propose to measure the risks by applying a privacy attack on these methods.}

Among privacy attacks, two types are particularly noteworthy: membership inference attacks~\cite{Shokri2017} aim to determine whether a specific data point was part of the training set of the target ML model. By querying the model and analyzing its behavior for different inputs, attackers can exploit the vulnerabilities caused by overfitting to infer the membership status of individual data points. Even more severe are reconstruction attacks~\cite{Carlini2019, Carlini2021}, which attempt to reconstruct the original data from the transformed or reduced-dimensional representations. These attacks typically assume the attacker has access to some data with the same distribution as the training set, knowledge of the model architecture, or statistical information about the population from which the training set is drawn.

Dimensionality reduction methods vary widely in their approaches, leading to differences in privacy leakage when their outputs are released. These differences stem from the varying degrees of randomization during training and the information preserved in the outputs. \textbf{This paper aims to compare privacy leakages from different dimensionality reduction methods using a reconstruction attack framework.} To achieve this, we propose a framework that includes a threat model and a reconstruction attack. Our framework is designed to be applicable across various dimensionality reduction methods. Additionally, we propose a defense strategy against the reconstruction attack, which, as a consequence, offers protection against weaker attacks.

\paragraph{\textbf{A summary of our contribution}} Our work presents the following contribution:
\begin{itemize}
    \item We propose a treat model and a reconstruction attack that can be applied to any dimensionality reduction method.
    \item Using this attack, we investigate and compare the privacy leakage from six popular dimensionality reduction methods. In our main study, we apply the attack on two specific datasets in order to identify underlying factors that affect the recontruction quality.
    \item  We investigate a simple additive noise mechanism and how much the noise's scale affects the reconstruction quality of each dimensionality reduction method.
\end{itemize}
As a result, our study reveals which dimensionality reduction methods are more resilient to privacy attacks, and which better retain the information after applying the defense mechanism.

\section{Background}

\subsection{Dimensionality reduction methods} \label{sec:drs}

\textbf{Setting.} Let $D = \{x_1,\ldots,x_n\} \in (\bR^{d})^n = \bR^{n \times d}$ be a dataset of $n$ data points in $\bR^d$. Given $d > 2$, a dimensionality reduction algorithm $M:\bR^{n \times d} \to \bR^{n \times 2}$ is a (possibly randomized) model that maps the $d$-dimensional dataset to $2$-dimensional one: $\theta = M(D)$. Here, $\theta$, referred to as a \emph{low-dimensional embedding} of $D$, can be reshaped as a vector of $2n$ parameters of $M$ fitted on $D$.

There are numerous dimensionality reduction methods. Among these, we shall focus on six methods that are often used on high-dimensional datasets~\cite{Nanga2021,Xia2022}. We give a quick overview of each dimensionality reduction technique.

% To be continued...
\begin{enumerate}
\item \textbf{Principal component analysis (PCA).}
PCA \cite{Pearson1901,Hotelling1933} is one of the most popular and widely used dimensionality reduction methods. It is used to reduce high-dimensional data into lower-dimensional representations by projecting the data onto directions with the highest variances. As a result, we obtain a lower-dimensional representation while preserving the variation in the dataset.

In particular, we reduce the data to two dimensions by projecting them to the first two principal components.

\item \textbf{Sparse random projection (SRP).}
\emph{Random projection} refers to a data reduction technique where each data point is projected onto a $r$-dimensional subspace ($r = 2$ in our case) via a random matrix $R$ of size $d\times r$, whose elements are typically sampled from a normal distribution: Denoting our original data by $D_{n\times d}$ and the random projection matrix by $R_{d\times r}$, the data in low dimension $\theta_{n\times r}$ is given by:
\begin{equation} \label{eq:PCA}
    \theta_{n\times r} = D_{n\times d}R_{d\times r}
\end{equation}
A well-known lemma by~\cite{Johnson1984} states that the projection approximately preserves the pairwise distances. Nonetheless, the complexity of the matrix multiplication is of order $O(ndr)$, which can be computationally expensive. To reduce the computation, sparse random projection~\cite{Achlioptas2001,Li2006} was proposed, in which the dense random matrix $R_{d\times r}$ is replaced by a sparse matrix $R^{\textsf{sparse}}_{d\times r}$, whose elements are:
\begin{equation} \label{eq:SRP}
% \[
R^{\textsf{sparse}}_{d\times r}[i,j] = \begin{cases}
        -\sqrt{\frac{d}{r}} \quad &\text{with probability } \frac{1}{2d} \\
        0 &\text{with probability } 1- \frac{1}{d} \\
        -\sqrt{\frac{d}{r}} &\text{with probability } \frac{1}{2d}
\end{cases}.
% \]
\end{equation}
As the average number of nonzero entries in each column of $R^{\textsf{sparse}}_{d\times r}$ is $d\cdot \frac1{d} = 1$, using this projection matrix results in a much faster computation of $O(nr)$ on average.~\cite{Li2006} shows that the embedding quality of SRP is comparable  to the classical random projection while being much more computationally efficient.

\item \textbf{Multidimensional scaling (MDS).}
MDS finds a low-dimensional embedding $\theta = \{ \theta_1,\ldots,\theta_n\}$ of data $D=\{x_1,\ldots,x_n\}$ through minimizing an objective known as the \emph{stress function}:
\begin{equation}\label{eq:stress}
\textsf{Stress}(\theta) = \sum_{1\leq i < j \leq n} (\lVert \theta_i -  \theta_j\rVert - \lVert x_i- x_j\rVert)^2.
\end{equation}
In other words, the stress function measures the discrepancies between the pairwise distances in the original space and those in the embedded space. There are many ways to minimize the function. In particular, we use an iterative steepest descent approach proposed by~\cite{Kruskal1964}.

\item \textbf{Isometric Mapping (Isomap).}

Isomap~\cite{Tenenbaum2000} was motivated by an observation that the Euclidean distance used in MDS fails to capture non-linear structure in high-dimensional data. For example, consider a simple data consisting of two points lying on opposite sides of a ball's surface. The Euclidean distance between these two points would not coincide with the surface distance from one point to the other. Isomap aims to embed data in a lower dimensional space while preserving the ``surface distance'' (referred to as \emph{geodesic distance}) in the high-dimensional space. This is done by performing the following steps:
\begin{enumerate}
\item Establish a graph $G$ with nodes $x_1,\ldots,x_n$. For each $x_i$, $x_j$ is connected to $x_i$ if is one of the $k$ nearest neighbors of $x_i$, and the edge is weighted by the distance $\lVert x_i - x_j \rVert$.
\item For all $i,j$, calculate the distance $d^G(x_i,x_j)$, given by the shortest path in $G$ from $x_i$ to $x_j$. Let $A^G$ be the distance matrix whose $(i,j)$-th entry is $d^G(x_i,x_j)$.
\item Apply MDS on the pairwise shortest path $d^G(x_i,x_j)$. It can be shown that the $r$-dimensional embedding is given by~\cite{Tenenbaum2000}:
% \[

\begin{equation} \label{eq:MDS}
\theta = \Lambda^{1/2}V^{\textsf{T}}
\end{equation}
% \]
where $\Lambda$ is the diagonal matrix of the top-$r$ eigenvalues and $V$ is the top-$r$ eigenvectors of $-\frac{1}{2}HA^GH^{\textsf{T}}$. Here, $H=I_n - \frac1{n}\mathbf{1}_n\mathbf{1}_n^{\textsf{T}}$ is the centering matrix (with $\mathbf{1}_n = (1,\ldots,1)^{\textsf{T}}$).
\end{enumerate}

\item \textbf{$t$-stochastic neighbor embedding ($t$-SNE).}
$t$-SNE~\cite{JMLR:v9:vandermaaten08a} approaches dimensionality reduction as an optimization problem, similar to MDS. The method begins by modeling the pairwise relationships between the original high-dimensional data points using a Gaussian kernel:

\begin{equation}\label{eq:tsne1}
% \[
p_{j\vert i} = \frac{e^{-\lVert x_i - x_j \rVert^2/2\sigma^2_i}}{\sum_{k \not= i}e^{-\lVert x_i - x_k \rVert^2/2\sigma^2_i}}, \quad\text{for }j\not= i, \qquad p_{i\vert i} = 0,
% \]
\end{equation}
where $\sigma_i$ is a data-dependent Gaussian bandwidth~\cite{JMLR:v9:vandermaaten08a}.

As $\sum_{i,j} p_{i\vert j} = n$, we define the symmetrized probability $p_{ij} = (p_{i\vert j} + p_{j\vert i})/2N$ which yields $\sum_{ij} p_{ij} = 1$. On the other hand, the pairwise distances between the embedded points $\theta = \{\theta_1,\ldots,\theta_n\}$ are modeled through the Student's $t$ kernel:

\begin{equation}\label{eq:tsne2}
% \[
q_{ij} = \frac{(1 + \lVert \theta_i - \theta_j\rVert^2)^{-1}}{\sum_k \sum_{l \not= k} (1 + \lVert \theta_k - \theta_l\rVert^2)^{-1}}
% \]
\end{equation}

With the discrete probability distributions $P = (p_{ij})_{1 \leq i,j \leq n}$ and $Q=(q_{ij})_{1 \leq i,j \leq n}$, the dissimilarities between the pairwise distances in the original and embedded spaces are measured through the Kullback-Leibler divergence between $P$ and $Q$:

\begin{equation}\label{eq:tsne3}
% \[
f_{t\text{-SNE}}(\theta) = \operatorname{KL}(P \Vert Q) = \sum_{i\not= j} p_{ij} \log \frac{p_{ij}}{q_{ij}}.
% \]
\end{equation}

A low-dimensional embedding can be obtained by minimizing $f_{t\text{-SNE}}(\theta)$ with respect to $\theta$. In the high-dimensional space, the Gaussian kernel used in $p_{ij}$ helps maintain the local structure of the data. Conversely, in the low-dimensional space, the heavy-tailed Student's $t$ kernel used in $q_{ij}$ allows for greater separation between dissimilar points, thereby preserving meaningful global structure in the embedding.

\item \textbf{Uniform manifold approximation and projection (UMAP).}
UMAP~\cite{McInnes2018} uses the same idea as $t$-SNE in that they model the pairwise distances in high-dimensional and low-dimensional spaces through discrete probabilities, with the following modifications: For each $x_i \in D$, let $\rho_i$ be the distance from $x_i$ to its nearest neighbor. With a specified number of nearest neighbors $k$, we define

\begin{equation}\label{eq:tsne4}
% \[
p_{i\vert j} = \begin{cases}
    e^{-(\lVert x_i - x_j \rVert - \rho_i)/\sigma_i} \quad &j \in \{ k \text{ nearest neighbors of }x_i \} \\
    0 &\text{otherwise}.
\end{cases}
% \]
\end{equation}
The symmetrized probability is defined as $p_{ij} = p_{i\vert j} + p_{j\vert i} - p_{i\vert j}p_{j\vert i}$. Note that $\sum_{ij} p_{ij}$ is not necessarily one. The pairwise distances in the low-dimensional space is modeled through:

\begin{equation}\label{eq:tsne5}
% \[
q_{ij} =  \frac{1}{1+a\lVert \theta_i - \theta_j \lVert^{2b}},
% \]
\end{equation}
which is similar to an unnormalized Student's $t$ kernel. Data-driven procedures to find suitable choices of $k$, $\sigma_i$, $a$ and $b$ are fully explained in~\cite{McInnes2018}.

Finally, with $P=(p_{ij})_{1\leq i,j \leq n}$ and $Q=(q_{ij})_{1\leq i,j \leq n}$, the optimization objective for UMAP is the cross-entropy loss:

\begin{equation}\label{eq:tsne6}
% \[
f_{\textsf{UMAP}}(\theta) = \operatorname{CE}(P,Q) = -\sum_{i\not= j} \left\{p_{ij}\log q_{ij} + (1-p_{ij}) \log (1-q_{ij}) \right\}.
% \]
\end{equation}
As with $t$-SNE, we minimize $f_{\textsf{UMAP}}(\theta)$ with respect to $\theta$ in order to find a suitable low-dimensional embedding of $D$.
\end{enumerate}

One key observation that will be brought up again in the discussion of our experiments is that these methods can be divided into two categories:
\begin{enumerate}
    \item \textbf{Deterministic methods.} These are PCA and Isomap. These methods always return the same output no matter how many times they are run on the same input.
    \item \textbf{Randomized methods.} These are SRP, MDS, $t$-SNE and UMAP. These methods produce different outputs on multiple runs as they are all initialized with random states.
\end{enumerate}

\subsection{Treat Model}
Our adversary model is from~\cite{Balle2022}, which is motivated by the treat model in differential privacy~\cite{Nasr2021}. We consider an adversary who do not have access to an original training point $x_i\in D$. We assume that this adversary has access to $A$'s output, as well as all $x_j$ for all $j \not= i$. With this information, their goal is to reconstruct the remaining training points $x_i$.

\begin{definition}[Informed adversary]\label{def:ads}
    Let $M$ be a dimensionality reduction algorithm, $D$ is a dataset and $\theta = M(D)$. Let $x_i$ be a training point in $D$. An \emph{informed adversary}'s goal is to reconstruct $x_i$ while having access to the following information:
    \begin{enumerate}
        \item $M$'s output $\theta$;
        \item $M$'s training algorithm;
        \item The \emph{known-member set}: $D_{-i} = D \setminus \{x_i\}$;
        \item A public dataset of $m$ data points $D' \in \bR^{m\times d}$ that is similar to $D$.
    \end{enumerate}
\end{definition}
In particular, the adversary might not have access to the intermediate calculations of $M$, such as gradients in iterative steps or matrices used in the computation.

While the framework of an informed adversary is unrealistic, it provides a robust framework for demonstrating the feasibility and extent of reconstruction attacks. By assuming an adversary with near-complete knowledge of the dataset, we can rigorously assess the maximum potential for data recovery through such attacks. In addition, this powerful adversarial model offers a stringent test for protective measures. Any defense mechanism capable of withstanding attacks from such a well-informed adversary would, by extension, prove effective against less knowledgeable and more realistic threats.

\subsection{Reconstruction Attack}
After retrieving $\theta, D_{-i}$ and $D'$ in Definition~\ref{def:ads}, the adversary introduces an attack $A$ that outputs a reconstruction $\hat{x}_i$ of $x_i$. Here, the quality of the reconstruction is measured by a distance function between $x_i$ and $\hat{x}_i$, which might be task-dependent. In this work, we will only consider the standard Euclidean distance $\lVert x_i - \hat{x}_i \rVert$.

The reconstruction attack is related to the attribute inference attack~\cite{Fredrikson2014, Fredrikson2015, Yeom2018, Zhang2020}, where some of $x_i$'s attributes are known to the adversary beforehand, and their goal is to estimate the remaining attributes. Reconstruction attacks can be seen as a generalization of attribute inference attacks, as the former can be used to accomplish the latter, with additional information from the public attributes.

While the original threat model of attribute inference attacks does not necessarily involve an informed adversary, we emphasize that our definition of adversary allows us to quantify the extent of reconstruction given near-complete information of the dataset. Consequently, any successful defense against an informed adversary would also be effective against any attribute inference attack, regardless of the adversary's level of prior knowledge.
\clearpage

\section{Our methods}
\begin{figure}[t]
    \centering
    \includegraphics[width=0.75\linewidth]{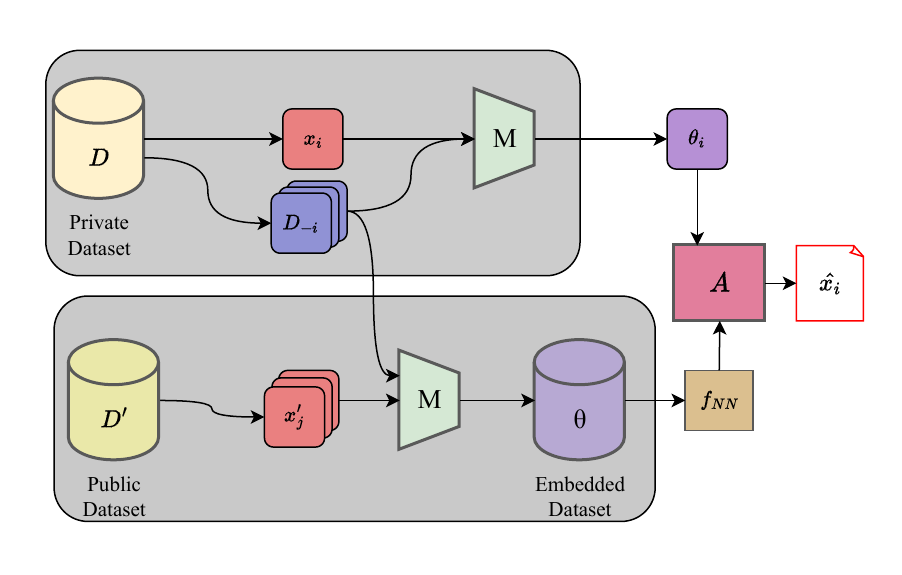}
    \caption{Overview of our reconstruction attack on dimensionality reduction methods.}
    \label{fig:enter-label}
\end{figure}

\subsection{An overview of the attack}

Our reconstruction attack is based on neural network-based attack proposed by~\cite{Balle2022}. An overview of the attack is displayed in Figure~\ref{fig:enter-label}.

First, we build a \emph{reconstruction network}, a neural network that takes the low-dimensional embedding $\theta$ as input, and then returns a reconstruction $\hat{x}_i$. The procedure to obtain a reconstruction network (\textsf{Train-Recon-Net} in Algorithm~\ref{alg:recon1}) consists of the following steps:

\begin{algorithm}[t]
\caption{Training reconstruction network}\label{alg:recon1}
\begin{algorithmic}[1]
    \Statex \hspace{-24pt} \textsf{Train-Recon-Net}$(D_{-i}, D', M, f_{NN})$
    \Require Known data points $D_{-i}$, public data $D'=\{x'_1,\ldots,x'_m\}$, target model's algorithm $M$, reconstruction neural network $f_{NN}: \mathbb{R}^{2n} \to \mathbb{R}^{d}$.
    \Ensure Reconstruction network $A$.
    \Statex {\color{blue}/* Create shadow training data */}
    \State Initialize \textsf{TrainSet} $=\{ \}$
    \For{$j = 1 \ldots, m$}
        \State $z_j \leftarrow D_{-i} \cup \{x'_j\}$ \Comment{Shadow input}
        \State $\theta_j \leftarrow M(z_j)$ \Comment{Shadow output}
        \State \textsf{TrainSet} $\leftarrow \textsf{TrainSet} \cup (\theta_j,x'_j)$
    \EndFor
    \Statex {\color{blue}/* Train reconstruction network */}
    \State $f \leftarrow$ Train $f_{NN}$ on \textsf{TrainSet}
    \State \textbf{Return:} $f$
\end{algorithmic}
\end{algorithm}

With the known data $D_{-i}$ and public data $D'= \{x'_1,\ldots,x'_m\}$. We create a \emph{shadow dataset} that consists of the following inputs and outputs:
\begin{itemize}
    \item \textbf{Input}: $\theta_j= M(z_j)$. Recall that $M$ is the dimensionality reduction model, and $z_j = D_{-i} \cup \{x'_j\}$. Here, $\theta_j \in \mathbb{R}^{2n}$ can be treated as a $2n$-dimensional input.
    \item \textbf{Output}: $x'_j$, which is a $d$-dimensional vector.
\end{itemize}
As a result, we obtain a dataset $\textsf{TrainSet}=\{(\theta_1,x'_1),\ldots,(\theta_m,x'_m)\}$. We then train a neural network $f_{NN}$ on \textsf{TrainSet} to obtain a reconstruction model. The architecture of $f_{NN}$ will be detailed in the next section.

\begin{algorithm}[t]
\caption{Reconstruction attack on dimensionality reduction models}\label{alg:recon2}
\begin{algorithmic}[1]
    \Statex \hspace{-24pt} $A(\theta, D_{-i}, D', M, f_{NN})$
    \Require Target model's output $\theta$, known data points $D_{-i}$, public data $D'=\{x'_1,\ldots,x'_m\}$, target model's algorithm $M$, reconstruction neural network $f_{NN}: \mathbb{R}^{2n} \to \mathbb{R}^{d}$.
    \Ensure Reconstruction $\hat{x}_i$ of the $i$-th data point.

    \State $f \leftarrow \textsf{Train-Recon-Net}(D_{-i}, D', M, f_{NN})$
    \State $\hat{x}_i \leftarrow f(\theta)$
    \State \textbf{Return:} $\hat{x}_i$
\end{algorithmic}
\end{algorithm}

As $\theta = M(D) = M(D_{-i} \cup \{x_i\})$, we then use the trained network to predict $x_i$ by taking $\theta$ as an input. Algorithm~\ref{alg:recon2} details the entire process of the reconstruction attack.

\subsection{Reconstruction network}
The architecture of $f_{NN}$ follows a standard encoder-decoder design.

The encoder (Table~\ref{tab:encoder}) takes an input of size $2n$ and produces an output of a smaller size. In our implementation, we use an output size of 80, though this can be adjusted as needed.

In contrast to previous work, the encoder is specifically designed to reconstruct $x_i$ from a low-dimensional embedding. The $2n$-dimensional input is first split into two components: the two-dimensional embedding of the target $x_i$, and the remaining $2n-2$ dimensions representing the embeddings of other data points. These split inputs are then processed by two separate feedforward layers:
\begin{itemize}
\item For the embedding of $x_i$, a layer with more than two hidden nodes is used in order to expand its representation.
\item For the embedding of the other points, a layer with fewer than $2n-2$ hidden nodes is used in order to compress the contextual information.
\end{itemize}
The outputs from these two layers are then concatenated to be passed to the decoder.

The decoder receives the encoder's output and generates a reconstruction matching the size of the original data in the high-dimensional space. Our design follows a standard decoder design for the corresponding output format. Specifically, if the output is one dimensional vector, then the decoder is a standard feedforward neural network. If the output is an image, then the decoder uses the stacked transposed convolutional layers as seen in U-Net-type networks~\cite{Ronneberger2015}. We will illustrate two examples of the decoders in our experiments on sample datasets.

\begin{table}[t]
\centering
\begin{tabular}{@{}lccc@{}}
\toprule
\textbf{Layer}  & \multicolumn{3}{c}{\textbf{Parameters}} \\ \midrule
Input           &             & $2n$         &              \\
Split           & $2$         & +          & $2n-2$         \\
Fully connected ($\times 2$) & $16$          & +          & $64$           \\
Concatenate     &             & $80$         &              \\ \bottomrule
\end{tabular}
\caption{Encoder of the reconstruction network.}
\label{tab:encoder}
\end{table}

\subsection{Previous work}

Reconstruction attacks based on neural networks were first proposed by~\cite{Balle2022}, who first introduce the notion of an informed adversary (Definition~\ref{def:ads}). In their framework, an adversary with access to a trained machine learning model attempts to reconstruct the model's training data. In particular, the authors propose an attack that uses a neural network, called a \emph{reconstructor network} (RecoNN), that outputs a reconstruction based on the model's parameters. In their work, the reconstruction attack is specifically employed to attack other neural networks. In addition, the authors demonstrate that differential privacy mechanisms can effectively mitigate this type of attack. Related attack is the attribute inference attack~\cite{Fredrikson2014,Fredrikson2015,Yeom2018,Zhang2020}, also referred to as model inversion attack~\cite{Zhang2020}. In this attacks, the adversary knows some of $x_i$'s features and, with access to the model, attempts to recover the remaining features. However, these attacks are not applicable to our setting, as our threat model assumes that the attacker has access to outputs of a dimensionality reduction method, not a neural network's parameters.

Another well-known attack is the membership inference attack~\cite{Shokri2017}, which aims to determine whether a particular data record was in the model's training set. This attack is performed with black-box queries, meaning that the adversary can only access the model's output on given inputs. To do so, they invented the shadow training technique, which comprises multiple ``shadow models'' of the target model, each of which is used to simulate members and non-members of the training set that most likely result in the target model's behavior. Then, the attack model is trained on simulated data to classify whether an input is a member or non-member of the training set.

To defend against these attacks, differential privacy~\cite{Dwork2006b,Dwork2006a} is considered a powerful privacy-preserving framework. A dataset query is differentially private if an adversary, even with knowledge of all but one data points, could hardly learn anything about the remaining data point from the query's output. In the context of dimensionality reduction, research has shown that the random projection can achieve differential privacy~\cite{Blocki2012,Kenthapadi2013}. A differentially private algorithm for PCA was first proposed by~\cite{Blum2005}, followed by a long line of research to improve the output quality, with notable contributions from~\cite{Chaudhuri13a,Hardt2013,Dwork2014,NIPS2014_729c6888} and~\cite{NEURIPS2022_c150ebe1}. Nonetheless, differential privacy analyses of other dimensionality reduction methods remain scarce, largely due to the complex nature of their outputs.

\begin{table}[t]
\footnotesize\centering
\begin{tabular}{@{}llll@{}}
\toprule
\textbf{Study}                               & \textbf{Attack}                                                        & \textbf{Datasets}                                                      & \textbf{Limitation}                                                                                 \\ \midrule
\cite{Shokri2017}              & \begin{tabular}[t]{@{}l@{}}Membership inference attack\end{tabular} & MNIST, CIFAR10                                                         & \begin{tabular}[t]{@{}l@{}}Not designed for \\ data reconstruction\end{tabular}                       \\
\\ \cite{10.1145/3325413.3329791} & Reconstruction attack                                                  & \begin{tabular}[t]{@{}l@{}}Google images, ImageNet\end{tabular}     & \begin{tabular}[t]{@{}l@{}}Intermediate outputs \\ (e.g. gradients) are \\ required\end{tabular}    \\
\\ \cite{Balle2022}               & Reconstruction attack                                                  & MNIST, CIFAR10                                                         & \begin{tabular}[t]{@{}l@{}}The target input is\\ treated as equally as\\ the remaining inputs\end{tabular} \\
\\ \cite{Zari2022}                & \begin{tabular}[t]{@{}l@{}}Membership inference attack\end{tabular} & \begin{tabular}[t]{@{}l@{}}MNIST, LFW, Adult, Census\end{tabular}   & \begin{tabular}[t]{@{}l@{}}Not designed for \\ data reconstruction\end{tabular}                       \\
\\ \cite{Kwatra2023}              & Reconstruction attack                                                  & \begin{tabular}[t]{@{}l@{}}Heart-scale, a9a, mushrooms\end{tabular} & \begin{tabular}[t]{@{}l@{}}Not designed to attack\\  low-dimensional \\  embeddings\end{tabular} \\
\\ Our approach                                 & Reconstruction attack                                                  & \begin{tabular}[t]{@{}l@{}}MNIST, NIH Chest X-ray\end{tabular}      & -                                                                                                   \\ \bottomrule
\end{tabular}
\caption{A summary of previous work on privacy attacks on machine learning models with their limitations in terms of attacks on dimensionality reduction methods.}\label{table:lit}
\end{table}

In terms of dimensionality reduction, recent works by~\cite{Zari2022} and~\cite{Kwatra2023} investigated membership inference and reconstruction attacks, but limited their scope to public releases of PCA’s principal components. In contrast, our work considers a different threat model where an attacker has access to the complete set of dimensionally-reduced data points produced by PCA—a format more commonly published in scientific papers than principal components. In addition, the published data points contain more information compared to principal components alone, leading to an easier attack.

 As summarized in Table~\ref{table:lit}, our work makes three novel contributions: (1) we present the first reconstruction attack that works across various dimensionality reduction techniques beyond PCA, (2) we propose a reconstruction neural network that produces higher quality reconstructions for the outputs of the deterministic methods (PCA and Isomap) compared to the previously proposed network by \cite{Balle2022}, and (3) we evaluate the effectiveness of additive noise as a defense mechanism against such attacks.

\paragraph[\textbf{How is attack on dimensionality reduction model different from previous work?}]{\textbf{How is attack on dimensionality reduction model different from previous work?}} Previous work on neural network-based reconstruction attacks have mainly focused on supervised learning models, using model parameters as input for reconstruction. Our method, in contrast, takes embedded data in low-dimensional space as input. This distinction results in the following key differences:
\begin{enumerate}
    \item \textbf{Input dimension.} Our reconstruction network takes low-dimensional embedded data as input with a dimension of $2n$. This is often smaller than the number of parameters in large neural networks.
    \item \textbf{Direct information extraction.} Dimensionality reduction models preserve a direct correspondence between input and output data. Specifically, the information of an input point $x_i$ is directly contained in the $i$-th row of the output. And the data points that are close to $x_i$ in the low-dimensional space are likely to be close to $x_i$ in the high-dimensional space as well. As a consequence, an informed adversary can gain significant information about $x_i$ with a complete knowledge of the remaining $n-1$ points. This is in contrast to supervised learning models, where individual data point information cannot be directly extracted from the model parameters. This observation motivates the modular design of the encoder in Table~\ref{tab:encoder}.
\end{enumerate}

\section{Experiments}
To evaluate our reconstruction attack on the six dimensionality reduction methods, we will apply the attack on two datasets: MNIST and NIH Chest X-ray datasets. In both experiments, we first randomly sample a known-member set $D_{-i}$ with size $\lvert D_{-i} \rvert \in \{9, 99, 199, 499, 999\}$. We formulate the reconstruction attack as a supervised learning problem, training a reconstruction network $f_{NN}$ on the dataset ${(\theta_1,x'_1),\ldots,(\theta_m,x'_m)}$, where each $\theta_j$ represents the low-dimensional embedding of $x'_j$ concatenated with set $D_{-i}$. We partition the remaining data into training, validation, and test sets, where each image represents the target data we would like to attack. After training the model on the training set, we evaluate the reconstruction attack's performance by computing the mean squared error between the reconstructed images and the actual images in the test set.

\textbf{Evaluation of attacks.}We evaluate the quality of our reconstructions quantitatively and qualitatively as follows:

\emph{Quantitative evaluation. }We compare the reconstructions (viewed as $d$-dimensional vectors) $\hat{x}_i^{recon} \in \mathbb{R}^d$ against the original images $x_i^{true} \in \mathbb{R}^d$ using the mean squared error (MSE):
\[  \text{MSE} = \frac{1}{dN} \sum_{i=1}^{N} \lVert x_i^{true}-\hat{x}_i^{recon} \rVert^2, \]
where $\lVert \cdot \rVert$ is the $d$-dimensional Euclidean norm

\emph{Qualitative evaluation. }We visualize the reconstructed images and qualitatively assess if they resemble the original images.

\textbf{Experimental environment details.} We have run both experiments on a Linux server running Ubuntu 22.04.3 LTS, with Intel(R) Xeon(R) CPU @ 2.00GHz processor with 29 GiB RAM, and NVIDIA P100 GPU with 16 GiB VRAM. The reconstruction networks are implemented and trained in PyTorch.

\subsection{MNIST dataset}
The MNIST dataset consists of grayscale images of handwritten digits ranging from 0 to 9. Their dimensions are $28\times 28$. The dataset has 60,000 images for training and 10,000 for testing. They are commonly used for training machine and deep learning models for classification and image processing tasks.

The selection of the number of data points in our approach is governed by computational resources. Consequently, the use of 5,000 uniformly random sampled data points is computationally feasible for training our neural network-based reconstruction models. Therefore, we took a random subset of 5,000 images from the (original) training set and divided them into three parts (1) a training set with 4,000 samples, (2) a validation set with 500 samples, and (3) a test set with 500 samples. From the remaining images, we randomly selected a set $D_{-i}$ with $\lvert D_{-i} \rvert \in \{9, 99, 199, 499, 999\}$ images to form the known-member set $D_{-i}$ (see Definition~\ref{def:ads}).

We prepare shadow datasets for the reconstruction network following Step $1-6$ outlined in Algorithm~\ref{alg:recon1}. Specifically, we apply each of the six dimensionality reduction methods (PCA, SRP, MDS, Isomap, $t$-SNE, and UMAP) to $x \cup D_{-i}$ for each $x$ in the training, validation, and test set. As a result, we obtained shadow training, validation, and test sets for the reconstruction network with the embedding as the input and the original image $x$ as the output.

\begin{table}[t]
    \centering
\begin{tabular}{@{}ll@{}}
\toprule
\textbf{Layer}        & \textbf{Parameters}                              \\ \midrule
Tranposed convolution & 512 filters of 4 $\times$ 4, strides 1, padding 0  \\
{\footnotesize \quad + Batchnorm + ReLU}    &                                                  \\
Tranposed convolution & 256 filters of 4 $\times$ 4, stride 2, padding 1   \\
{\footnotesize \quad + Batchnorm + ReLU}    &                                                  \\
Tranposed convolution & 128 filters of 4 $\times$ 4, strides 2, padding 1 \\
{\footnotesize \quad + Batchnorm + ReLU}    &                                                  \\
Tranposed convolution & 64 filters of 4 $\times$ 4, strides 2, padding 1 \\
{\footnotesize \quad + Batchnorm + ReLU}    &                                                  \\
Tranposed Convolution & 1 filters of 1 $\times$ 1, strides 1, padding 2 \\
ReLU                  & $28 \times 28$ units                                          \\ \bottomrule
\end{tabular}
    \caption{Decoder of the reconstruction network for the MNIST dataset.}
    \label{tab:decoder}
\end{table}

In the training stage, the shadow training set was split into minibatches, each with a size of 64. The architectures of the reconstruction network's encoder and decoder are described in Table~\ref{tab:encoder} and~\ref{tab:decoder}, respectively. In particular, as the output of the network is a 2D image, the decoder consists of multiple transposed convolutional layers. We also applied batch normalization, and ReLU as the activation function. Our loss function is a combined MSE loss and L1 loss, with Adam~\cite{Kingma2015} as the optimizer. Moreover, even though we defined the epoch at 500 and the learning rate at $1\times10^{-5}$, early stopping was very essential because it can prevent models from overfitting.

The data preparation and training procedure were repeated for every dimensionality reduction method and every known-member set $D_{-i}$ with size $\lvert D_{-i} \rvert \in \{9, 99, 199, 499, 999\}$. In addition, to account for variation in the known-member sets, we repeated each combination five times. We then reported the average and the standard deviation of the MSE on the shadow test set.

\begin{figure}[t]
    \centering
    \includegraphics[width=0.8\textwidth]{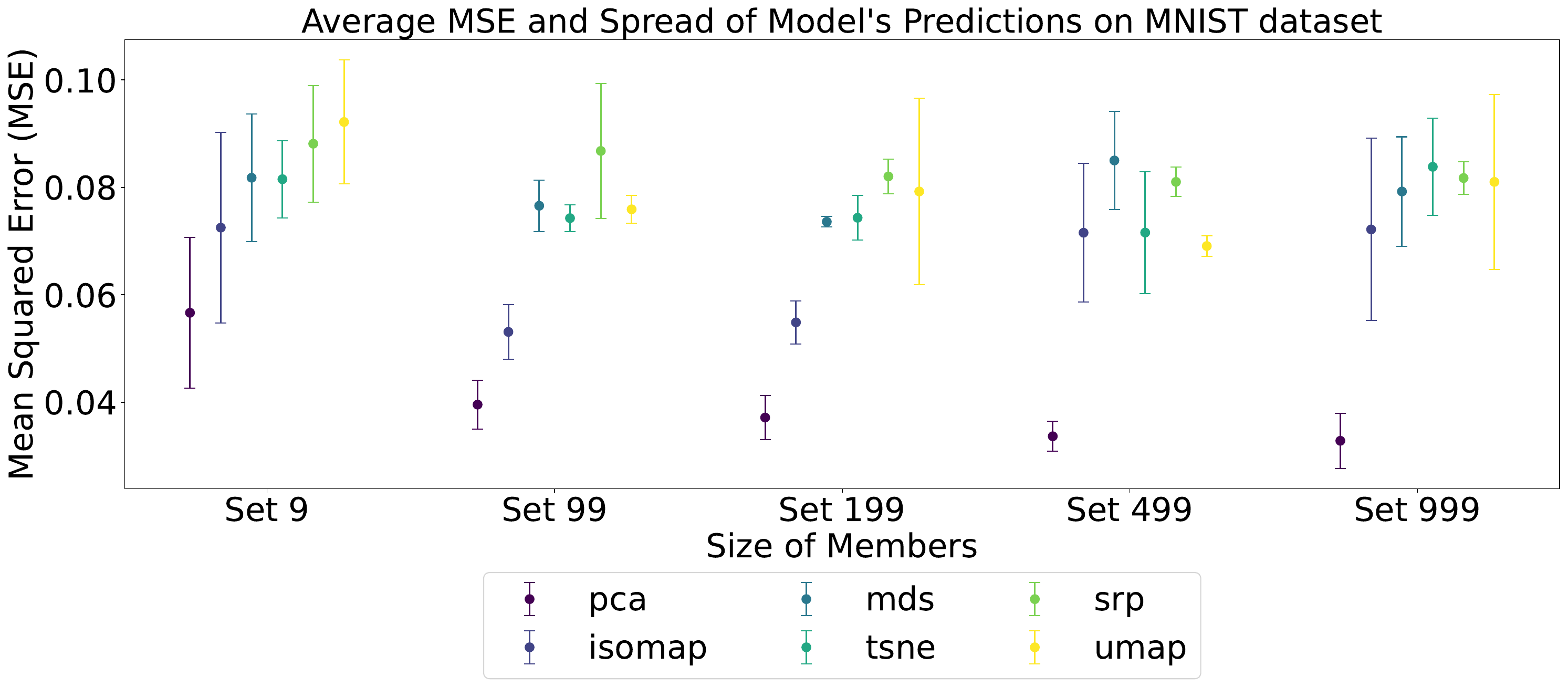}
    \caption{Average MSE with one standard deviation of the reconstructions on six dimensionality reduction models trained on the MNIST dataset, with the number of the known members ranging from 9 to 999.}
    \label{fig:average_mse_mnist}
\end{figure}

Figure~\ref{fig:average_mse_mnist} presents the reconstruction MSEs for the six dimensionality reduction methods across the five known-member set sizes. The results reveal that PCA is the most vulnerable method, indicated by its comparatively small MSEs. Moreover, PCA's MSE shows a declining trend as the known-member set size increases, indicating that the quality of reconstructions improves when the adversary has access to more data points.

For the other models, attacks on Isomap perform well on small known-member set sizes. However, as the set size increases to 499 or 999, attacks on the Isomap and the other models besides PCA becomes more difficult as indicated by their relatively large MSEs. By comparing the averages, we notice that MDS is consistently the third most vulnerable model, while SRP is the least vulnerable model across all set sizes.

\begin{figure}[t]
    \centering
    \includegraphics[width=0.5\textwidth]{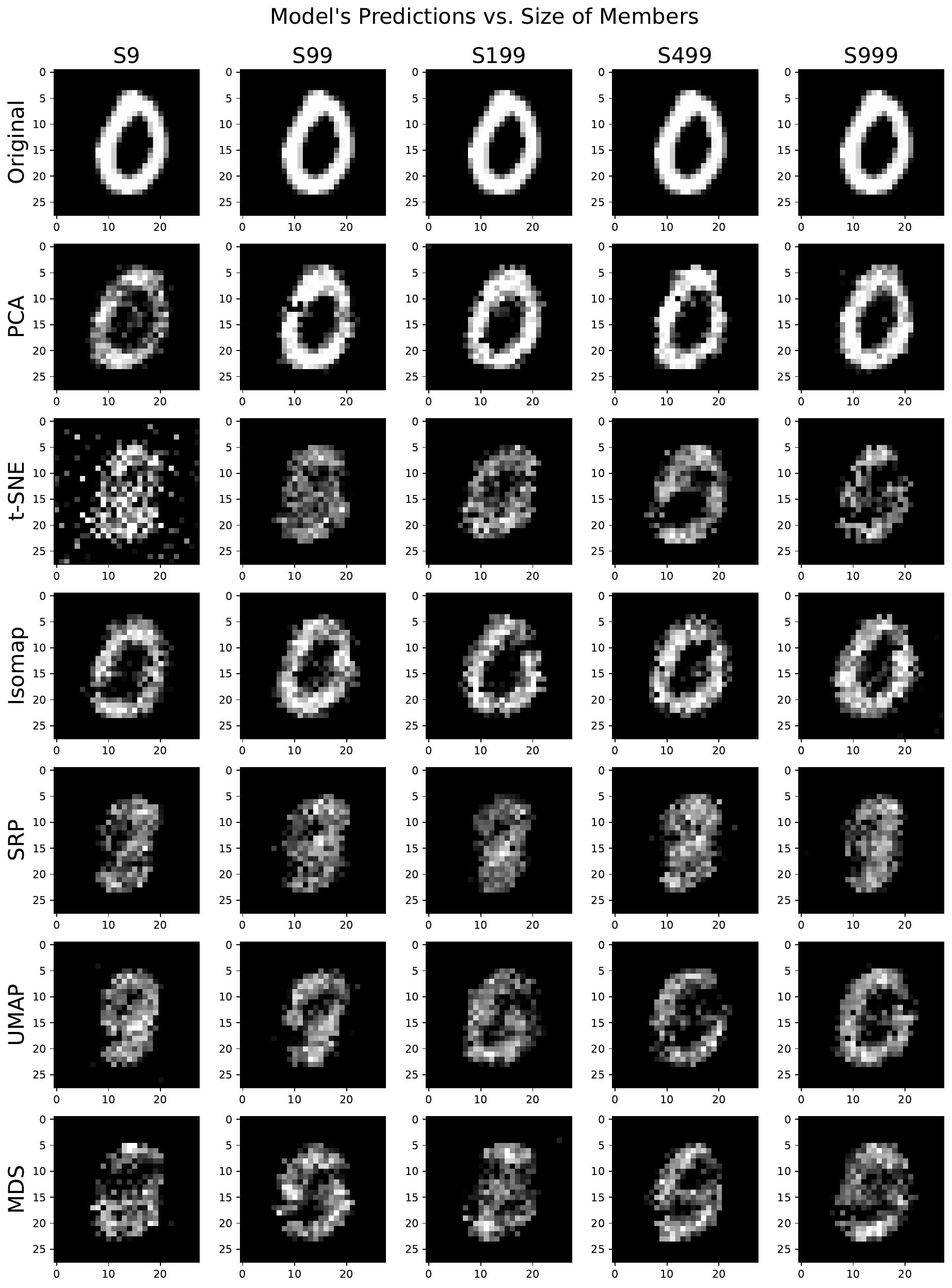}
    \caption{Examples of outputs from reconstruction attacks on six dimensionality reduction models trained on the MNIST dataset, with the number of known members varying from $9$ to $999$.}
    \label{fig:model-pred}
\end{figure}

Examples of an image reconstructions are displayed in Figure~\ref{fig:model-pred}. We can see that the reconstructions from attacks on PCA and Isomap closely resemble the original image, aligning with the lower MSE values discussed earlier. For the other methods, attacks on $t$-SNE, MDS, and UMAP show moderate success when the known-member set size is 499 or 999. In contrast, attacks on SRP is the least successful among all methods tested, as their reconstructions barely resemble the original image.

We can see that the MSE of the PCA and Isomap, which are methods with non-randomized initializations, are easier to attack than SRP, MDS, $t$-SNE, and UMAP, which use random initialization.

We have also compared the performance of our network with \cite{Balle2022}'s network for MNIST reconstruction, which consists of two dense layers with 1000 nodes each and the ReLU activation function. The reconstruction quality of our network and \cite{Balle2022}'s network, measured in terms of the MSE against the original images, are shown in Table~\ref{table:comparenn}. In particular, the table shows that, for the deterministic methods (PCA and Isomap), our network significantly improves in the reconstruction quality over the previously proposed network.

\subsection{NIH Chest X-ray dataset}
For our second experiment, we considered a more complex dataset. We considered the NIH Chest X-ray dataset~\cite{Wang2017}, which consists of 112,120 frontal-view X-ray images of 30,805 unique patients. We first reduced the image size from $1024\times1024$ to $64\times64$. We randomly selected a subset of 5,000 images from the dataset, which was then separated into 4,000 images for training, 500 images for validation, and 500 images for testing.

\begin{table}[t]
\centering \footnotesize
\begin{tabular}{rcccccc}
\toprule
Network      & PCA & SRP & MDS & Isomap & $t$-SNE & UMAP \\ \midrule
\cite{Balle2022}  & $ 0.056\pm 0.004$    &  $ 0.079 \pm 0.008$   & $ 0.075 \pm 0.006$   &  $ 0.068 \pm 0.003$      &   $ 0.077\pm 0.001$      &   $ 0.079 \pm 0.001$   \\
Ours &  $ \mathbf{0.037}\pm 0.004$   &  $ 0.082 \pm 0.003$   &  $ 0.074\pm 0.001$ &  $ \mathbf{0.055} \pm  0.004$ & $ \mathbf{0.074} \pm 0.004$      &  $ 0.079\pm 0.017$    \\ \bottomrule
\end{tabular}
\caption{Comparison between our reconstruction network and the network proposed by \cite{Balle2022} in terms of image reconstruction quality measured in MSE. The attacks were performed with the size of the known-member set $\lvert D_i \rvert = 199$.}
\label{table:comparenn}
\end{table}

Following the same data preparation procedure for MNIST, we constructed shadow training, validation, and test sets from the original dataset. We changed our model's architecture a little bit to fit the size of an output image (as detailed in Table~\ref{tab:decoder2}). However, all hyperparameters were still the same as MNIST.

\begin{table}[t]
    \centering
\begin{tabular}{@{}ll@{}}
\toprule
\textbf{Layer}        & \textbf{Parameters}                              \\ \midrule
Tranposed convolution & 512 filters of 4 $\times$ 4, strides 1, padding 0  \\
{\footnotesize \quad + Batchnorm + ReLU}    &                                                  \\
Tranposed convolution & 1024 filters of 4 $\times$ 4, stride 2, padding 1   \\
{\footnotesize \quad + Batchnorm + ReLU + Dropout}    &                                                  \\
Tranposed convolution & 512 filters of 4 $\times$ 4, strides 2, padding 1 \\
{\footnotesize \quad + Batchnorm + ReLU}    &                                                  \\
Tranposed convolution & 256 filters of 4 $\times$ 4, strides 2, padding 1 \\
{\footnotesize \quad + Batchnorm + ReLU}    &                                                  \\
Tranposed Convolution & 1 filters of 4 $\times$ 4, strides 2, padding 1 \\
ReLU                  & $64 \times 64$ units                                          \\ \bottomrule
\end{tabular}
    \caption{Decoder of the reconstruction network for the NIH Chest X-ray dataset.}
    \label{tab:decoder2}
\end{table}

\begin{figure}[t]
    \centering
    \includegraphics[width=0.8\textwidth]{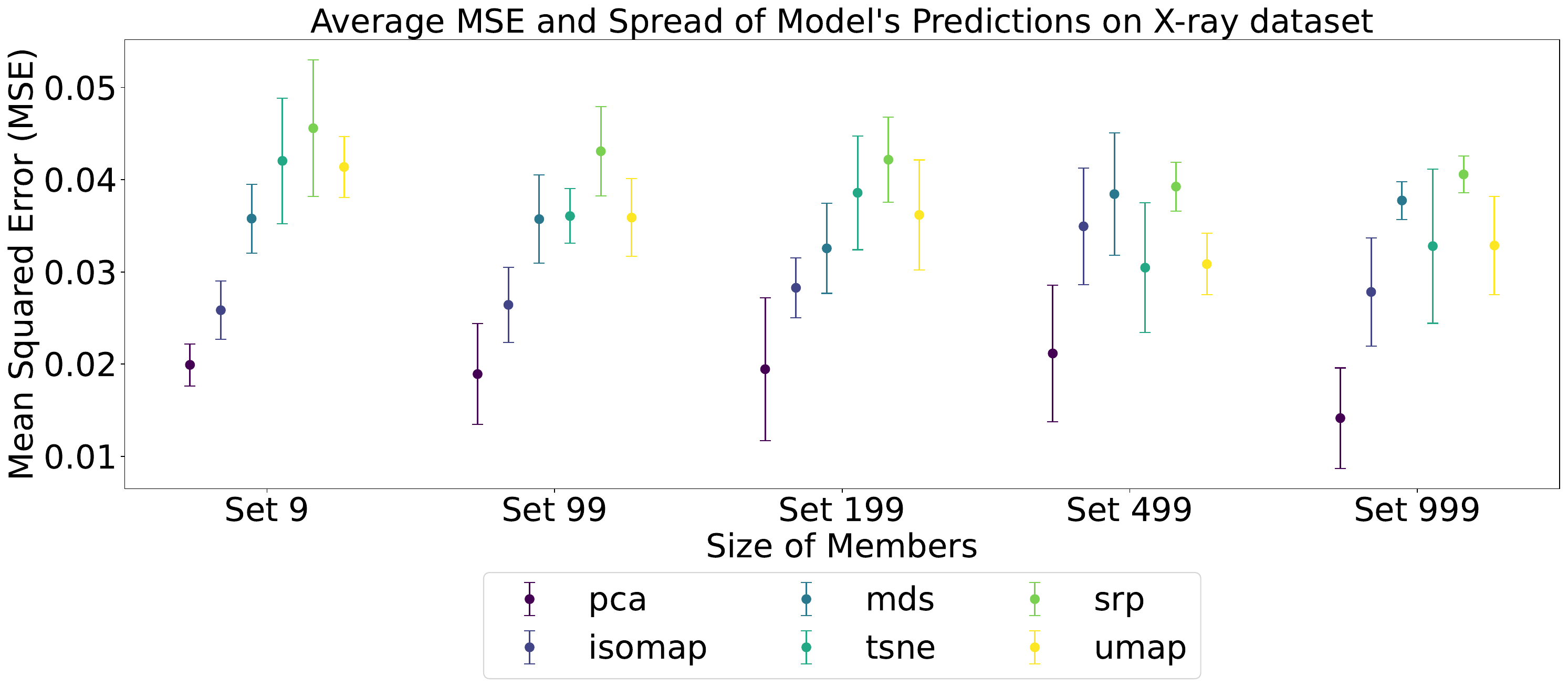}
    \caption{Average MSE with one standard deviation of the reconstructions on six dimensionality reduction models trained on the NIH Chest X-ray dataset, with the number of the known members ranging from 9 to 999.}
    \label{fig:average_mse_med}
\end{figure}

We repeated the data preparation and model training five times to account for the variation in the known-member sets. Figure~\ref{fig:average_mse_med} presents the average MSE along with one standard deviation for each method and each known-member set size. Consistent with our findings from the MNIST dataset, these results indicate that PCA remains the most vulnerable dimensionality reduction model. However, unlike the MNIST results, we observe no clear declining trend in the average MSE as the known-member set size increases. In addition, Isomap is the second most vulnerable method (except at $ \lvert D_{-i} \rvert=499$), and SRP is the least vulnerable method.

\begin{figure}[t]
    \centering
    \includegraphics[width=0.5\textwidth]{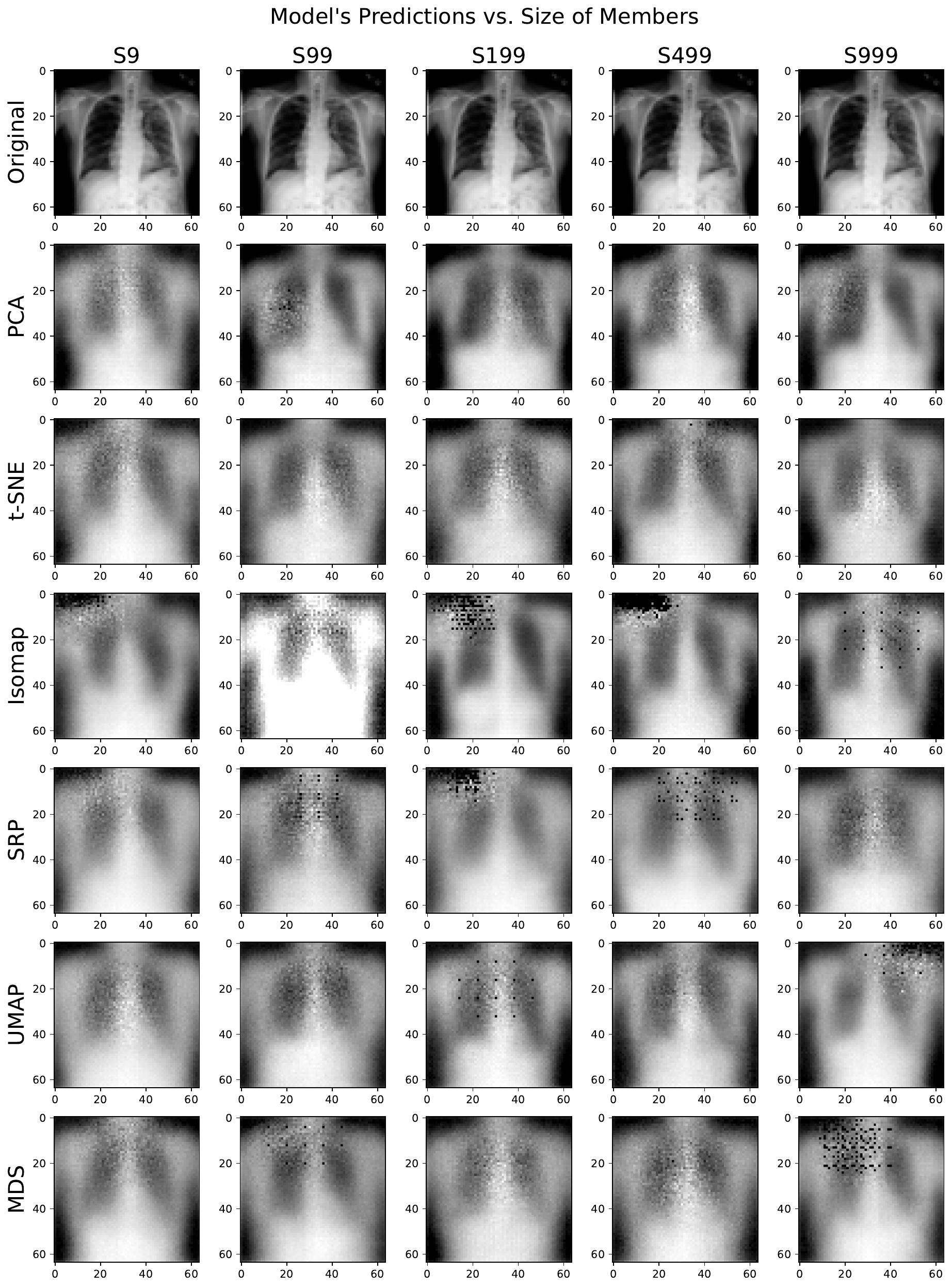}
    \caption{Examples of outputs from reconstruction attacks on six dimensionality reduction models trained on the NIH Chest X-ray dataset, with the number of known members varying from $9$ to $999$.}
    \label{fig:pred_vs_members_med}
\end{figure}

Figure~\ref{fig:pred_vs_members_med} displays examples of X-ray image reconstructions from our attacks. Unlike the MNIST results, these reconstructions show notably poorer quality across all methods and known-member set sizes, though they still bear some resemblance to the original images. Nonetheless, we can still notice differences in the reconstruction quality among the methods. Attacks on PCA consistently produce the best reconstructions, as indicated by darker chest areas and the absence of artifacts. In contrast, attacks on SRP is the least successful, as they produce the fuzziest images compared to the other methods.

\section{Discussion}

Here, we summarize our key findings from the two experiments:

\begin{itemize}
    \item The attack is most effective against PCA and least effective against SRP across both datasets, as demonstrated both quantitatively by the MSE plots and qualitatively through the reconstruction examples.

    \item The attack is effective on non-randomized algorithms and ineffective on randomized algorithms.

    The deterministic methods (PCA and Isomap) not only produce the same results when applied on the same dataset, but also show only slight changes in output when a single data point is modified. As a result, it is relatively easy for the reconstruction network to identify patterns in the low-dimensional representations and accurately recreate the original high-dimensional data.

    The randomized methods (SRP, MDS, $t$-SNE and UMAP), on the other hand, start with random initializations that can end up with totally different outputs in different runs, even when fitting on the same dataset. This results in diverse patterns in the shadow dataset that are particularly difficult for our reconstruction network to learn.

    Our results indicate that when data privacy is a primary concern in the publication of low-dimensional embeddings, randomized algorithms are preferable, as their inherent variability makes them more resistant to the reconstruction attack.
\end{itemize}

\begin{itemize}
    \item Our results show that our reconstruction network proposed in Table \ref{tab:encoder} and \ref{tab:decoder} provide higher reconstruction quality than those from the network proposed by \cite{Balle2022}. This is because our network is designed to amplify the signal from the data point that we wish to attack, with $1/4$ of the nodes in the hidden layers dedicated to this particular point. In contrast, the standard network in \cite{Balle2022} was designed to process each data point equally. As a result, our network is able to transform the low-dimensional embedding of the target point to the original image.

    \item Our results show the most significant improvement when increasing the size of the known-member set from $ \lvert D_{-i} \rvert=9$ to $ \lvert D_{-i} \rvert=99$. However, for some models, further increases in $D_{-i}$ lead to decreased attack performance. This counterintuitive effect occurs because the input dimension of the reconstruction network ($2n$) grows with $D_{-i}$ (where $ \lvert D_{-i} \rvert = n-1$). As the input dimension increases, the network requires more complexity and a larger number of shadow training samples to effectively learn from such high-dimensional data.

    This observation suggests that the attack can potentially become stronger as $D_{-i}$ increases, but this can only be achieved with a larger public dataset, a more complex network architecture and more computation resources.

\end{itemize}

\section{Defense against the reconstruction attack}

To defend against reconstruction attacks, we investigate an additive noise mechanism. This approach involves adding independent noise with dimensions corresponding to a $28\times 28$ matrix, sampled from a normal distribution, to the high-dimensional data prior to dimensionality reduction. To assess how the scale of noise affects reconstruction quality, we apply this technique to the MNIST dataset. Specifically, we add normally distributed noise to each image $x$ as follows:
\begin{equation}
x+\varepsilon, \qquad \varepsilon \sim \mathcal{N}(\mathbf{0},\,\sigma^{2}I_{784})\,.
\end{equation}
where $784$ is the dimension of the MNIST image $x$, $\sigma \in \{2, 4, 8, 16, 32\}$, and $I_{784}$ is the $784\times 784$ identity matrix. To quantitatively evaluate the noisy outputs, we perform the following procedure:
\begin{itemize}
    \item Each of the dimensionality reduction methods is applied on the real and noisy images.
    \item We perform the attack on both the methods’ outputs that correspond to the real and noisy images.
    \item The attack outputs reconstructions $x^{true}\in \mathbb{R}^{784}$ and $x^{noisy} \in \mathbb{R}^{784}$, with which we calculate the reconstruction error using the MSE:
    \begin{equation}\label{eq:mse}
    \text{MSE}(x^{true}, x^{noisy})  = \frac{1}{N}  \sum_{i=1}^{N} \lVert x^{true}_i -  x^{noisy}_i\rVert^2,
    \end{equation}
    where $\lVert \cdot \rVert$ is the $784$-dimensional Euclidean norm.
\end{itemize}
Thus, the additive noise mechanism is ineffective if the MSE is small, that is, if the noisy reconstruction is close to the real one.

\begin{figure}[t]
    \centering
    \includegraphics[width=0.8\textwidth]{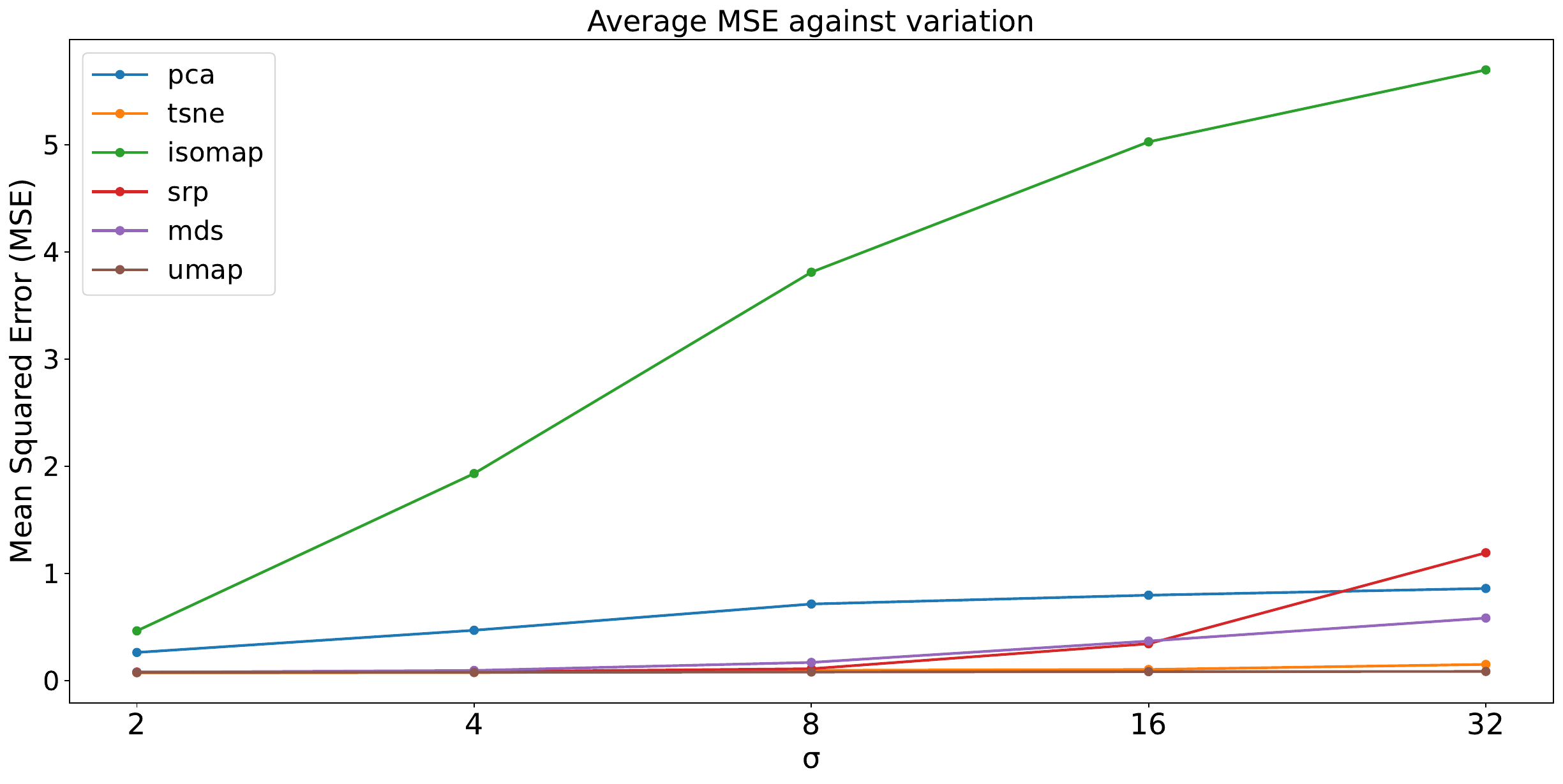}
    \caption{Average MSE on each scale of noises for every dimensionality reduction methods.}
    \label{fig:average_mse_noises}
\end{figure}

We feed noisy images into various dimensionality reduction methods, whose outputs are then passed on to our reconstruction networks that were trained on non-noisy data. Figure~\ref{fig:average_mse_noises} illustrates the MSEs of the reconstruction with $ \lvert D_{-i} \rvert=99$ for each dimensionality reduction model.
The results demonstrate that the additive noise mechanism is highly effective for certain reduction methods. Notably, Isomap appears to be the most significantly affected by the added noise, as indicated by relatively large MSEs compared to other models.

\begin{figure}[t]
    \centering
    \includegraphics[width=0.5\textwidth]{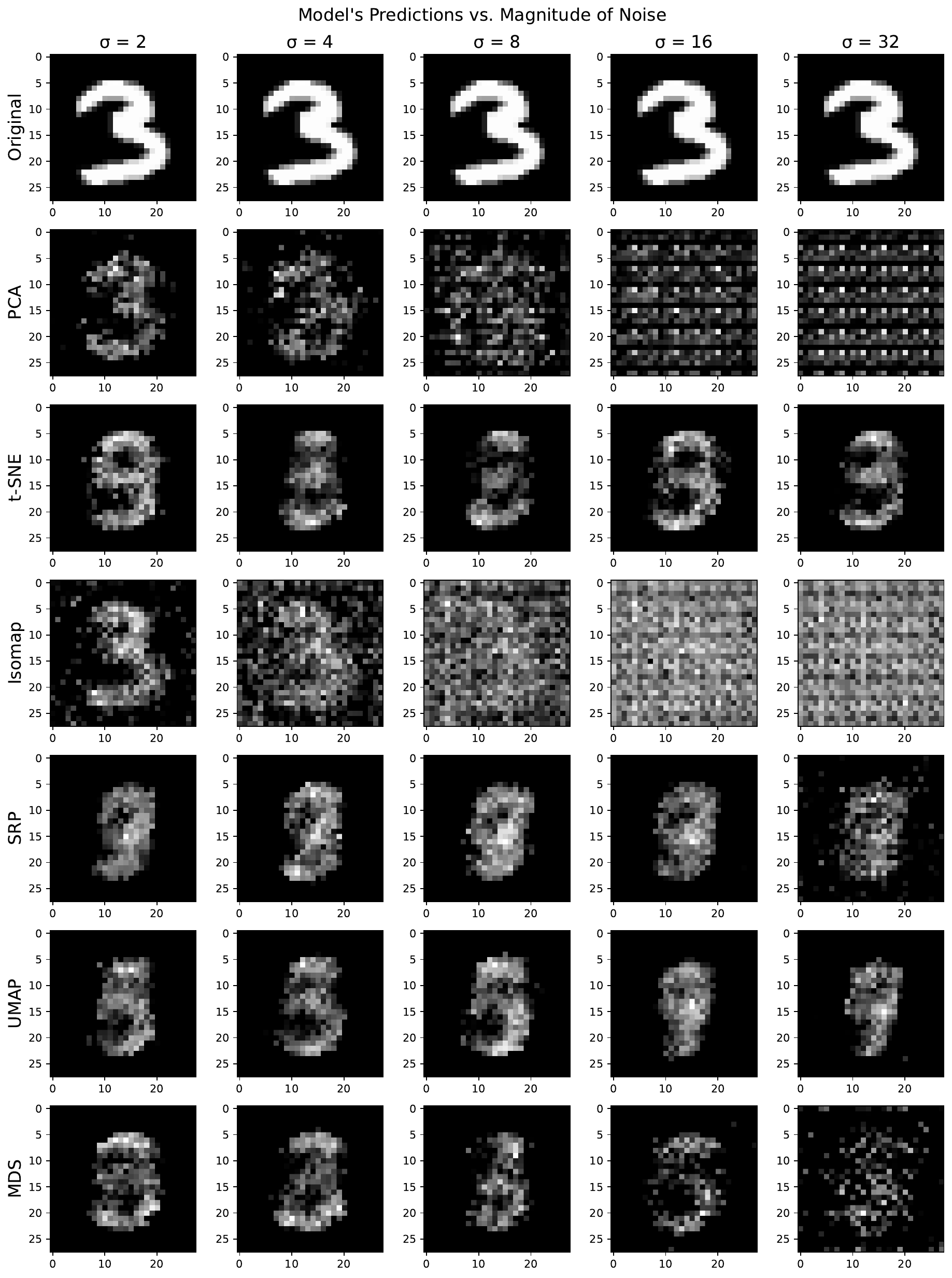}
    \caption{Outputs of reconstruction attacks on embeddings of a noisy image in MNIST.}
    \label{fig:pred_data_with_noise}
\end{figure}

Figure~\ref{fig:pred_data_with_noise} provides examples of reconstructions from noisy images. Interestingly, while only Isomap shows high MSE values, both PCA and Isomap reconstructions become fully distorted at high noise levels, suggesting that privacy is preserved in these two models' outputs after adding noises to the original image. In particular, the reconstruction of Isomap with large noises consists mostly of high pixel values, which is consistent with the high MSEs in Figure~\ref{fig:average_mse_noises}. We also observe that the reconstructions for the randomized methods somewhat resemble a digit even with large noises, indicating that the reconstruction networks have learned sufficiently diverse outputs from these methods that it is able to output digit-like images even from noisy lower-dimensional embeddings.

As explained above, the additive noise mechanism totally distorts the outputs of PCA and Isomap, which have non-randomized initializations. For the other methods, namely SRP, MDS, $t$-SNE, and UMAP, the reconstructions still somewhat resemble those of the original images, indicating that our attack has learned sufficient patterns of these models’ outputs even with random initialization.

\section{Conclusion}

Dimensionality reduction methods can help improve data interpretation and reduce model training costs. However, while these methods do not directly reveal the original data, they may inadvertently leak sensitive information. In this study, we propose a reconstruction attack in the context of an informed adversary to assess how much individual information is leaked from the published outputs of various dimensionality reduction methods.

We propose to use a neural network to reconstruct the original data. Our proposed network is specifically designed to exploit the order-preservation nature of the dimensionality reduction models and separate the low-dimensional embedding of an individual from the rest. In our experiments, we perform the reconstruction attack on six dimensionality reduction models on the MNIST and NIH Chest X-ray datasets. The results empirically indicate that the reconstruction attack is relatively effective on the deterministic algorithms, namely PCA and Isomap. We also measure the reconstruction quality in terms of the MSE between the actual images and the reconstructions from the outputs. The deterministic algorithms, namely PCA and Isomap, closely resemble the original images. In contrast, methods that utilize random initialization, namely SRP, MDS, $t$-SNE, and UMAP, show greater resilience to these attacks. In particular, based on both quantitative metrics and qualitative analysis, SRP demonstrates the highest resilience among the evaluated models. We also find that The attack's effectiveness increases with the adversary's knowledge of data points. However, this improvement comes at a cost, as it requires a larger public dataset, a more complex model architecture, and more computational resources.

We also suggest an additive noise mechanism to defend against the reconstruction attack. Our experiment shows that the reconstructions for PCA and Isomap become highly distorted when added with large noises, implying that the mechanism is effective for these two methods. The mechanism is also effective for the methods with random initializations: Though the reconstructions show some recognizable features, they do not resemble the original images. This observation aligns with our quantitative analysis (with the MSE between the actual and reconstructed images) of the reconstruction qualities.

\newcommand{\etalchar}[1]{$^{#1}$}

\end{document}